\documentstyle [12pt]{article}

\begin{document}

\begin{titlepage}

\title{Proposed experimental tests of the
Bell-Kochen-Specker theorem\footnote{Phys. Rev. Lett. 
{\bf 80}, 1797 (1998).}}

\author{Ad\'{a}n Cabello\thanks{Electronic 
address: fite1z1@sis.ucm.es}\\
{\em Departamento de F\'{\i}sica Aplicada,}\\
{\em Universidad de Sevilla, 41012 Sevilla, Spain.}\\
\\
Guillermo Garc\'{\i}a-Alcaine\thanks{Electronic 
address: fite114@sis.ucm.es}\\
{\em Departamento de F\'{\i}sica Te\'{o}rica,}\\
{\em Universidad Complutense, 28040 Madrid, Spain.}}

\date{\today}

\maketitle


\begin{abstract}
For a two-particle two-state system, 
sets of compatible propositions exist for which 
quantum mechanics 
and noncontextual hidden-variable theories make conflicting 
predictions for every individual system whatever its 
quantum state. This permits a simple all-or-nothing 
state-independent experimental verification 
of the Bell-Kochen-Specker theorem.\\
\\
PACS numbers: 03.65.Bz

\end{abstract}

\end{titlepage}

\begin{sloppypar}
There are two main theorems on the impossibility of
hidden variables in quantum mechanics (QM). The most general
is the Bell-Kochen-Specker (BKS) theorem \cite{Bell66,KS67}
which excludes  noncontextual hidden-variable (NCHV) theories 
(i.\ e., those in which the values of the physical observables
are the same whatever the experimental context in which they appear).
The other is Bell's theorem \cite{Bell64} which discards 
local hidden variables of the kind considered by
Einstein, Podolsky, and Rosen \cite{EPR35}.
Both theorems are mathematical statements which, as such, do not 
require any real experiment to be proved or disproved.
Only if we want to investigate how nature behaves do we require 
actual experiments.
There is a wide range of experiments which show 
that nature violates Bell's inequalities \cite{Aspectetal}.
However, no empirical disproof of NCHV theories 
has yet been exhibited \cite{Santos88}. 
This situation can be explained by comparing the proofs of both theorems.
Bell's inequalities \cite{Bell64} involve statistical magnitudes 
which can be calculated from measurements carried out
in different subensembles of pairs. In contrast,
the proofs of the BKS theorem \cite{Bell66,KS67,varios} 
refer to a single individual system but 
involve noncompatible observables that cannot be measured in the
same individual system. 
On the other hand, while Bell's inequalities 
are verified by any local hidden-variable theory 
independently of any QM assumptions, 
the proofs of the BKS theorem refer to
NCHV theories that share some properties with QM.
In this sense, the proofs of the BKS theorem are not entirely 
independent of the formal structure of QM.
For these reasons, one could think that 
``the whole notion of an experimental test of [B]KS misses the point'' 
\cite{Merminpc}.
\end{sloppypar}

In this paper we will show a situation, the first to our knowledge, 
in which NCHV
theories, {\em without} any call to the formal structure of
QM, make conflicting predictions with those of QM 
for every individual system and whatever its quantum state.
These predictions can be tested by a joint measurement
of {\em one} set of compatible propositions. 

We propose the following situation. Consider an individual 
system of two
spin-$\frac{1}{2}$ particles (or any other two-particle two-state
system) initially prepared in an 
unspecified state. Suppose that a NCHV theory can
describe that system. Noncontextuality here will mean
that this hidden-variable theory satisfies the following 
two assumptions: 

(i) Any one-particle observable (for a two-state system) can be 
assumed to have a definite value. 
This is a reasonable assumption for any NCHV theory since Gleason's 
theorem \cite{Gleason57} is not 
valid for systems described
by a Hilbert space of dimension two, and since the possibility of NCHV
for these systems was explicitly proved by Bell \cite{Bell66} and by
Kochen and Specker \cite{KS67}.
In particular, we will assume that the observables $A:=\sigma _z^{(1)}$, 
$B:=\sigma _z^{(2)}$,  
$a:=\sigma _x^{(1)}$, and $b:=\sigma _x^{(2)}$ (the spin 
components in units of ${\hbar}/{2}$ 
in the $z$ and $x$ directions for particles one and two)
have predefined noncontextual values either $+1$ or $-1$. 
We will denote these values as $v(A)$, $v(B)$, $v(a)$, and $v(b)$.
Then, considering the values of these four observables, 
$2^4$ different ``states'' could exist
(for instance, one possible ``state'' is 
$v(A)=-v(B)=-v(a)=v(b)=+1$).

(ii) The value of a two-particle observable 
which is a product of one-particle observables such as 
$AB$ (or $Ab$, etc) is
\begin{equation}
v(AB):=v(A)\,v(B)\,.
\label{product}
\end{equation}
Note that $A$ and $B$ are not only 
compatible observables but refer
to two different particles \cite{FineTeller78}.
Definition (\ref{product}) is a consequence of noncontextuality 
since one particular way of measuring the
observable $AB$ is by measuring separately $A$ and $B$ and multiplying
their results; but, in a NCHV theory, 
$v(AB)$ must be the same whatever the experimental context in which
it appears.

Now we will show some predictions derived 
from these two assumptions. For that 
purpose consider the following four propositions:
\begin{equation}
{P}_1:= ``AB=1\;\;\;\mbox{and}\;\;\;ab=1"\,,
\label{proposition1}
\end{equation}
\begin{equation}
{P}_2:=``AB=-1\;\;\;\mbox{and}\;\;\;ab=-1"\,,
\label{proposition2}
\end{equation}
\begin{equation}
{P}_3:=``Ab=1\;\;\;\mbox{and}\;\;\;aB=1"\,,
\label{proposition3}
\end{equation}
\begin{equation}
{P}_4:=``Ab=-1\;\;\;\mbox{and}\;\;\;aB=-1"\,.
\label{proposition4}
\end{equation}
Proposition ${P}_1$ has the value 1 ({\em true}) if the two-particle
observables $AB$ {\em and} $ab$ have values $+1$, 
and the value 0 ({\em false}) otherwise, etc. In a NCHV theory, 
the $\{{P}_i\}$ have
predefined values related, using assumption (ii), to those of $A$, $B$, 
$a$, and $b$. For instance,
$v({P}_1) = 1$ if $v(A) = v(B)$ {\em and} $v(a) = v(b)$, and zero
otherwise. 

As can be easily seen from the study of all the possible states 
of this NCHV theory, some predictions can be made:

NCHV1.---The propositions ${P}_1$, ${P}_2$, ${P}_3$, ${P}_4$ are {\em not}
mutually {\em exclusive}: Two of them can be simultaneously true 
[for instance, $v({P}_1)=v({P}_3)=1$ 
in the state $v(A)=v(B)=v(a)=v(b)=+1$].

NCHV2.---${P}_1$, ${P}_2$, ${P}_3$, ${P}_4$ are {\em not exhaustive}:
All of them can be simultaneously false 
[for instance, $v({P}_1)=v({P}_2)=v({P}_3)=v({P}_4)=0$ 
in the state $v(A)=v(B)=v(a)=-v(b)=+1$].

Indeed, checking all the possible states, NCHV1 and NCHV2
can be summarized as follows:

NCHV3.---In a NCHV theory, the values of  
${P}_1$, ${P}_2$, ${P}_3$, and ${P}_4$ in a joint 
measurement would be either 
4 zeros---all the propositions are false---, 
or 2 ones and 2 zeros---2 propositions are true and 2 
are \mbox{false---}.

Note that the predictions NCHV1, NCHV2, and NCHV3 are 
entirely independent of the formal structure of QM.

What are the corresponding quantum predictions?
First, let us see the quantum representatives of propositions
${P}_1$, ${P}_2$, ${P}_3$, and ${P}_4$.
If $\hat {A}$, $\hat {B}$, $\hat {a}$, and $\hat {b}$
denote the self-adjoint operators representing the observables 
$A$, $B$, $a$, and $b$, the proposition
${P}_i$ is represented by the projector 
$\hat P_i:= 
\left| {\psi _i} \right\rangle \left\langle {\psi _i} \right|$,
where $\left\{ {\left| {\psi _i} \right\rangle } \right\}$
are the states 
defined by the following eigenvalue equations \cite{states}: 
\begin{equation}
\hat A\otimes \hat B\;\left| {\psi _1} \right\rangle =
\left| {\psi _1} \right\rangle\,,\;\;\;
\hat a\otimes \hat b\;\left| {\psi _1} \right\rangle =
\left| {\psi _1} \right\rangle\,,
\label{projector1}
\end{equation} 
\begin{equation}
\hat A\otimes \hat B\;\left| {\psi _2} \right\rangle = 
-\left| {\psi _2} \right\rangle\,,\;\;\;
\hat a\otimes \hat b\;\left| {\psi _2} \right\rangle = 
-\left| {\psi _2} \right\rangle\,,
\label{projector2}
\end{equation}
\begin{equation}
\hat A\otimes \hat b\;\left| {\psi _3} \right\rangle =
\left| {\psi _3} \right\rangle\,,\;\;\;
\hat a\otimes \hat B\;\left| {\psi _3} \right\rangle =
\left| {\psi _3} \right\rangle\,,
\label{projector3}
\end{equation} 
\begin{equation}
\hat A\otimes \hat b\;\left| {\psi _4} \right\rangle =
 -\left| {\psi _4} \right\rangle\,,\;\;\;
\hat a\otimes \hat B\;\left| {\psi _4} \right\rangle = 
-\left| {\psi _4} \right\rangle\,.
\label{projector4}
\end{equation} 

As can be easily seen, the projectors $\hat P_1$, 
$\hat P_2$, $\hat P_3$, and $\hat P_4$ 
are mutually orthogonal, 
\begin{equation}
\hat{P}_i\, \hat{P}_j = 0\;\;\;{\em {\em if}}\;\;\; i\ne j\,.
\label{orthogonality}
\end{equation}
Therefore, according to QM:

QM1.---The propositions ${P}_1$, ${P}_2$, ${P}_3$, and ${P}_4$
are mutually exclusive: Two of them cannot 
be simultaneously true.

Moreover, it can be checked that the projectors $\hat P_1$, 
$\hat P_2$, $\hat P_3$, and $\hat P_4$ form 
a resolution of the identity, i.e.,
\begin{equation}
\hat{P}_1 + \hat{P}_2 + \hat{P}_3 + \hat{P}_4 =  \hat{1}\,.
\label{resolution}
\end{equation}
Therefore, according to QM:

QM2.---${P}_1$, ${P}_2$, ${P}_3$, and ${P}_4$ are exhaustive: Not 
all of them can be simultaneously false.

Indeed, from the mathematical properties (\ref{orthogonality}) 
and (\ref{resolution}) follows a third physical
prediction which includes QM1 and QM2:

QM3.---According to QM, in any joint measurement of 
${P}_1$, ${P}_2$, ${P}_3$, and ${P}_4$ in the same individual system, 
one and only one
of the propositions will be true and the other three will be false, 
whatever the preparation of the state.

Clearly, NCHVi and QMi are conflicting physical predictions. 
The situation at this point is
similar to that which appears between Bell's inequalities and QM: 
We have two theories with contradictory predictions.
Now we have to propose an experiment to check how nature behaves.

How could a joint measurement of 
${P}_1$, ${P}_2$, ${P}_3$, and ${P}_4$ be possible? 
Until now we have assumed that the propositions
${P}_1$, ${P}_2$, ${P}_3$, and ${P}_4$ are {\em compatible}. 
This remains to be justified. Of course, we have seen that
the projectors $\hat P_1$, 
$\hat P_2$, $\hat P_3$, and $\hat P_4$ commute, and it is a
generally accepted assumption of QM that commuting 
operators correspond to compatible observables. The reason for
this assumption is that, if there is a set of
pairwise commuting self-adjoint operators, then there exists 
a nontrivial {\em maximal}---nondegenerate--- operator $\hat H$ commuting 
with all $\hat P_i$, such that $\hat P_i =f_i(\hat H)$ \cite{vonNeumann31}. 
However, this justification hinges on the existence of a physical 
observable $H$ which corresponds to the operator $\hat H$. 
In our case, such operator can be
\begin{equation}
\hat H=\sum\limits_{i=1}^4 {c_i}\,\hat P_i\,,
\label{opmax}
\end{equation}
where the $\{c_i\}$ are arbitrary distinct real numbers. 
Then, it is easily checked that
\begin{equation}
\hat P_i=\prod\limits_{j\ne i} {{{\hat H-c_j \hat 1} \over {c_i-c_j}}}\,.
\end{equation}
Optical observables corresponding to operators of the form 
(\ref{opmax}) for two-particle systems have been proposed and 
actual experimental results are expected to be presented soon \cite{ZZH97}. 
On the other hand, the proposals \cite{Moussa97etal} 
for experiments designed to measure 
the {\em Bell operator} 
\cite{BMR92} used for quantum teleportation \cite{BBCJPW93} can
be modified to measure operators of the form (\ref{opmax}) \cite{bellstates}.

In summary, we have showed that there are situations in nature in which 
NCHV theories, without any call to the formal structure of QM, make
conflicting predictions with those of QM for every individual system 
whatever its quantum state. An experimental test of these
predictions requires the measurement of a particular set of 
compatible propositions. Optical versions of experiments 
related with these propositions
have been proposed for other purposes, and actual 
experimental results based on these proposals 
are expected to be presented soon.
\\

The authors wish to thank Asher Peres for his comments and suggestions, 
and David Mermin for his observations and criticisms; both
have been essential in the writing of this Letter.
We also acknowledge comments by Ignacio Cirac and Emilio Santos. 
One of us (A. C.) thanks Harvey Brown, Gonzalo Garc\'{\i}a de Polavieja, 
and Erik Sj\"{o}qvist for useful discussions, and for
their hospitality at Oxford.

\pagebreak


\begin{thebibliography}{99}
\bibitem{states} We can write down explicitly these states on 
the basis of common eigenstates of 
$\sigma _z^{(1)}$ and $\sigma _z^{(2)}$:  
$\left| {\psi _1} \right\rangle ={1 \over {\sqrt 2}}
\left( {\left| {++} \right\rangle +\left| {--} \right\rangle } \right)$, 
$\left| {\psi _2} \right\rangle ={1 \over {\sqrt 2}}
\left( {\left| {+-} \right\rangle -\left| {-+} \right\rangle } \right)$, 
$\left| {\psi _3} \right\rangle ={1 \over 2}
\left( {\left| {++} \right\rangle +\left| {+-} \right\rangle +
\left| {-+} \right\rangle -\left| {--} \right\rangle } \right)$,
$\left| {\psi _4} \right\rangle ={1 \over 2}
\left( {-\left| {++} \right\rangle +\left| {+-} \right\rangle +
\left| {-+} \right\rangle +\left| {--} \right\rangle } \right)$.
\bibitem{bellstates} Indeed, 
the states $\left| {\psi _i} \right\rangle$ 
appearing in Eqs. (\ref{projector1})-(\ref{projector4}) 
are almost the same as those of the
Bell operator basis, 
$\left| {\Phi ^\pm } \right\rangle $ and 
$\left| {\Psi ^\pm } \right\rangle $ (Ref. \cite{BMR92}), namely
$\left| {\psi _1} \right\rangle =\left| {\Phi ^+} \right\rangle$,
$\left| {\psi _2} \right\rangle =\left| {\Psi ^-} \right\rangle$,
$\left| {\psi _3} \right\rangle ={1 \over {\sqrt 2}}\left( {\left| {\Psi ^+} 
\right\rangle +\left| {\Phi ^-} \right\rangle } \right)$, and
$\left| {\psi _4} \right\rangle ={1 \over {\sqrt 2}}\left( {\left| {\Psi ^+} 
\right\rangle -\left| {\Phi ^-} \right\rangle } \right)$.
\end{thebibliography}
\end{document}